\documentclass[a4paper,fleqn,usenatbib]{mnras}

\usepackage{mathptmx}

\usepackage[T1]{fontenc}
\usepackage{ae,aecompl}

\usepackage{graphicx}
\usepackage{amsmath}	
\usepackage{amssymb}	
\usepackage{mathrsfs}
\usepackage{url}
\usepackage{bm}
\usepackage{booktabs}
\usepackage{multicol}
\usepackage{comment}
\usepackage[font=small,skip=0pt]{caption}

\newcommand{\msun}{\ensuremath{\,\textrm{M}_{\odot}}}

\newcommand{\A}{\ensuremath{\alpha}}

\title[The impact of ECSNe on DNSs]{The impact of electron-capture supernovae on merging double neutron stars} 

\author[Giacobbo \& Mapelli]{
Nicola Giacobbo,$^{1,2,3,4}$\thanks{E-mail: giacobbo.nicola@gmail.com}
Michela Mapelli,$^{2,3,4}$
\\
$^{1}$Dipartimento di Fisica e Astronomia ``G. Galilei'',
    Universit\`a di Padova, vicolo dell'Osservatorio 3, I--35122 \\
$^{2}$INAF, Osservatorio Astronomico di Padova, vicolo dell'Osservatorio 5, I--35122 Padova, Italy.\\
$^{3}$INFN, Milano Bicocca, Piazza della Scienza 3, I--20126, Milano, Italy\\
$^{4}$Institut f\"ur  Astro- und Teilchenphysik, Universit\"at Innsbruck, Technikerstrasse 25/8, A--6020, Innsbruck, Austria\\ 
}

\date{Accepted XXX. Received YYY; in original form ZZZ}

\pubyear{2018}

\begin{document}
\label{firstpage}
\pagerange{\pageref{firstpage}--\pageref{lastpage}}
\maketitle

\begin{abstract}
  Natal kicks are one of the most debated issues about double neutron star (DNS) formation. Several observational and theoretical results suggest that some DNSs have formed with low natal kicks ($\lesssim{}50$~km~s$^{-1}$), which might be attributed to electron-capture supernovae (ECSNe). We investigate the impact of ECSNe on the formation of DNSs by means of population synthesis simulations. In particular, we assume a Maxwellian velocity distribution for the natal kick induced by ECSNe with one dimensional root-mean-square $\sigma_{\rm ECSN} = 0,7,15,26,265$~km~s$^{-1}$. The total number of DNSs scales inversely with $\sigma_{\rm ECSN}$ and the number of DNS mergers is higher for relatively low kicks. This effect is particularly strong if we assume low efficiency of common-envelope ejection (described by the parameter $\alpha=1$), while it is only mild for high efficiency of common-envelope ejection ($\alpha{}=5$).  In most simulations, more than 50 per cent of the progenitors of merging DNSs  undergo at least one ECSN and the ECSN is almost always the first SN occurring in the binary system. 
  Finally, we have considered the extreme case in which all neutron stars receive a low natal kick ($\lesssim{}50$~km~s$^{-1}$). In this case, the number of DNSs increases by a factor of ten and the percentage of merging DNSs which went through an ECSN is significantly suppressed ($<40$ per cent).
\end{abstract}

\begin{keywords}
methods: numerical -- gravitational waves -- binaries: general -- stars: neutron
\end{keywords}



\section{Introduction}
GW170817, the first detection of a merger between two neutron stars \citep[NSs,][]{Abbott2017b}, marked the beginning of multi-messenger astronomy. For the first time, electromagnetic emission accompanying the gravitational wave (GW) event was observed \citep{Abbott2017c}, ranging from gamma rays (e.g. \citealt{Abbott2017c,Goldstein2017,Savchenko2017}) to X-rays (e.g. \citealt{Margutti2017}), to optical, near-infrared (e.g. \citealt{Coulter2017,Soares-Santos2017,Chornock2017,Cowperthwaite2017,Nicholl2017,Pian2017}) and radio  wavelengths (e.g. \citealt{Alexander2017}).

The formation of merging double NSs (DNSs) like GW170817 is still matter of	 debate: understanding this process would provide crucial insights for both stellar evolution and GW astrophysics. Merging DNSs are expected to form either from the evolution of isolated close binaries (e.g. \citealt{Flannery1975,Bethe1998,Belczynski2002,Voss2003,Dewi2003,Podsiadlowski2004,Dewi2005,Andrews2015,Tauris2017,Chruslinska2017,Kruckow2018,Vigna2018,Giacobbo2018b,Mapelli2018,Mapelli2018b}) or through dynamical interactions in star clusters (e.g. \citealt{Grindlay2006,East2012,Lee2010,Ziosi2014}).

Many uncertainties still affect both formation channels. In particular, one of the most debated and also one of the most important  physical ingredients for the formation of DNSs 
is the magnitude of the natal kick imparted by the supernova (SN) explosion to the newborn NS \citep{Janka2012}.

From a study on the proper motion of 233 young isolated pulsars, \citet{Hobbs2005} estimated that their velocity distribution follows a Maxwellian curve with a one dimensional root-mean-square (1D rms) velocity $\sigma=265$~km~s$^{-1}$ and an average natal kick speed of $\sim 420$~km~s$^{-1}$. On the other hand, there is increasing evidence that some NSs form with a significantly smaller natal kick.

Several studies \citep{Cordes1998,Arzoumanian2002,Brisken2003,Schwab2010,Verbunt2017} claim that the velocity distribution proposed by \citet{Hobbs2005} underestimates the number of pulsars with a low velocity and suggest that the natal kick distribution of NSs is better represented by a bimodal velocity distribution. This bimodal distribution might result from two different mechanisms of NS formation. For instance, two out of nine accurate pulsar velocities computed by \citet{Brisken2002} are smaller than $40$~km~s$^{-1}$. Moreover, \citet{Pfahl2002} study a new class of high-mass X-ray binaries with small eccentricities and long orbital periods, which imply a low natal kick velocity ($\lesssim 50$~km~s$^{-1}$) for the newborn NSs. Similarly, \cite{Knigge2011} show that Be X-ray binaries could be divided in two sub-populations: one with short ($\sim{}10$ s) and one with long ($\sim{}200$ s) spin period. The two populations are characterized also by different orbital period and eccentricity distributions, indicative of two natal kick distributions. 
Even considerations about the orbital elements of some Galactic DNSs suggest that a low natal kick is required \citep{Heuvel2007,Beniamini2016}.

 It has been proposed that NSs with a low natal kick come from electron-capture SNe (ECSNe, \citealt{Miyaji1980,Nomoto1984,Nomoto1987,Heuvel2007}), a more rapid and less energetic process with respect to iron core-collapse SNe (CCSNe, \citealt{Dessart2006,Kitaura2006}). In ECSN explosions, the asymmetries are more difficult to develop and the newborn NS receives a lower kick \citep{Dessart2006,Jones2013,Schwab2015,Gessner2018}.

Low natal kicks might occur not only in ECSNe, but in all low-mass progenitors ($\lesssim{}10$ M$_\odot$), because of their steep density profile at the edge of the core, which allows for rapid acceleration of the SN shock wave. The shock is revived on a shorter time scale than in more massive progenitors, and therefore there is less time for large-scale asymmetries (which would result in a larger kick) to develop (see e.g. \citealt{Mueller2016}).

   Alternatively, the low kick of some DNSs could be explained also by the fact that they come from ultra-stripped SNe, i.e. from the SN explosion of a naked Helium star that was stripped by its compact companion \citep{Tauris2013,Tauris2015,Tauris2017}. In this case, the natal kick is thought to be lower because of the low mass of the ejecta. 

In this paper, we use our new population-synthesis code {\sc MOBSE} \citep{Giacobbo2018}, to investigate the impact of ECSNe and low natal kicks on the formation of merging DNSs. 
 We show that ECSNe are an important channel for the formation of DNSs, if they are associated to low natal kicks. Moreover, we discuss the extreme case in which all NSs receive a small kick, regardless of the SN process.



\begin{table}
	\begin{center}
		\caption{Definition of the simulation sets.\label{tab:ecsim}}
		\begin{tabular}{cccc}
			\toprule
			     ID & $\sigma_{\rm{ECSN}}$ & $\sigma_{\rm{CCSN}}$ & $\alpha$ \\
                            			 & [km~s$^{-1}$] & [km~s$^{-1}$] & \\
			\midrule
			EC0\A1 & 0.0  & 265.0  & 1 \\
			EC7\A1 & 7.0  & 265.0  & 1\\
			EC15\A1 & 15.0  & 265.0  & 1\\
			EC26\A1 & 26.0  & 265.0  & 1\\
			EC265\A1 & 265.0  & 265.0  & 1\vspace{0.2cm}\\ 
                        EC0\A5 & 0.0  & 265.0  & 5 \\
			EC7\A5 & 7.0  & 265.0  & 5\\
			EC15\A5 & 15.0  & 265.0  & 5\\
			EC26\A5 & 26.0  & 265.0  & 5\\
			EC265\A5 & 265.0  & 265.0  & 5\vspace{0.2cm}\\ 
			CC15\A1 & 15.0  & 15.0  & 1\\ 
			CC15\A5 & 15.0  & 15.0  & 5\\ 
			\bottomrule	
		\end{tabular}
	\end{center}
	    {\small Column 1: simulation name; column 2-3: 1D rms of the Maxwellian natal kick distribution for ECSNe and CCSNe, respectively (see sec.~\ref{sec:2.2}); column 4: values of $\alpha$ in the CE formalism (see sec.~\ref{sec:2.3}). Simulations CC15$\alpha{}1$ and CC15$\alpha{}5$ are the same as we already presented in \cite{Giacobbo2018b}.}	
\end{table}

\section{Methods}
\label{sec:2}
      {\sc MOBSE} is an updated version of the {\sc BSE} code \citep{Hurley2000,Hurley2002}. Here we summarize the main characteristics of {\sc MOBSE} and we describe the new features we have added to it for this work. A more detailed discussion of {\sc MOBSE} can be found in \cite{Giacobbo2018} and in \cite{Mapelli2017}. In this paper, we adopt the version of {\sc MOBSE} described as {\sc MOBSE1} in \cite{Giacobbo2018}.

      The main differences between {\sc MOBSE} and {\sc BSE} are the treatment of stellar winds of massive stars and the prescriptions for SN explosions. Stellar winds of O and B-type stars are implemented in {\sc MOBSE} as described by \cite{Vink2001}, while the mass loss of Wolf-Rayet (WR) stars is implemented following \cite{Belczynski2010}. Finally, the mass loss of luminous blue variable (LBV) stars is described as
		\begin{equation}\label{eq:LBV}
		  \dot{M} = 10^{-4}\,{}f_{\rm LBV}\,{}\left(\frac{Z}{\text Z_{\odot}}\right)^{\beta{}}\,{} {\rm M}_{\odot}\,{}{\rm yr}^{-1},
		\end{equation}
            
                where $f_{\rm LBV}=1.5$ \citep{Belczynski2010} and $Z$ is the metallicity.
                  
    In {\sc MOBSE}, all massive hot massive stars (O, B, WR and LBV stars) lose mass according to $\dot{M}\propto{}Z^{\beta}$, where $\beta{}$ is defined as \citep{Chen2015}

\begin{equation}
\label{eq:scaling}
\beta = \begin{cases}
  0.85 & \rm{if}~~~ \Gamma_{\rm e} < 2/3 \cr
  2.45 - 2.40 \,{} \Gamma_{\rm e} & \rm{if}~~~ 2/3 \leq \Gamma_e \leq 1~ \cr
  0.05\,{} & \rm{if}~~~ \Gamma_{\rm e}>1,
\end{cases}
\end{equation}
where $\Gamma_{\rm e}$ is the electron-scattering Eddington ratio, expressed as (see eq.~8 of \citealt{Graefener2011}):
\begin{equation}\label{eq:gamma}
\log{\Gamma_e}=-4.813+\log{(1+X_{\rm H})}+\log{(L/L_\odot)}-\log{(M/M_\odot)}.
\end{equation}
In equation~\ref{eq:gamma}, $X_{\rm H}$ is the Hydrogen fraction, $L$ is the star luminosity and $M$ is the star mass.


      The new prescriptions for core-collapse SNe (CCSNe) in {\sc MOBSE} include the rapid and the delayed SN model described  by \cite{Fryer2012} (see also \citealt{Spera2015}). The rapid SN model is adopted for the simulations presented in this paper, because it allows us to reproduce the remnant mass gap between $\sim{}2$ M$_\odot$ and $\sim{}5$ M$_\odot$ \citep{Ozel2010,Farr2011}. Pair-instability and pulsational pair-instability SNe are also implemented in {\sc MOBSE} using the fitting formulas by \cite{Spera2017}.

      Finally, we have also updated the prescriptions for core radii following \cite{Hall2014}, we have extended the mass range up to 150 M$_\odot$ \citep{Mapelli2016}, and we have revised the treatment of Hertzsprung-gap (HG) donors in common envelope (CE): HG donors are assumed to always merge with their companions if they enter a CE phase.

      For this work, we have added several updates to the description of ECSNe and natal kicks in {\sc MOBSE}, as we describe in the following sections. 
	\begin{figure*}
		\includegraphics[scale=0.4]{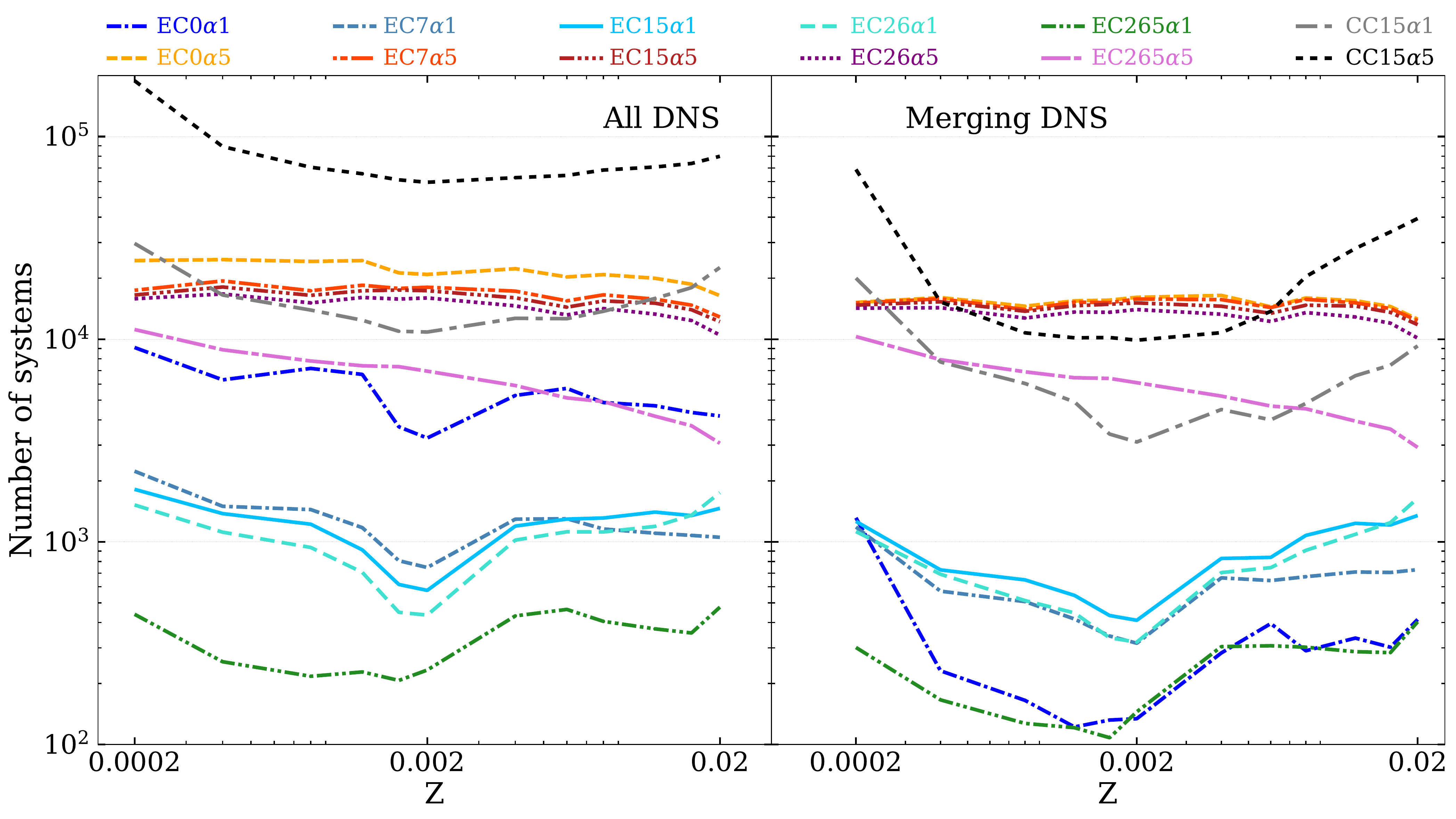}
		\caption{Impact of different kick velocities for ECSNe on the number of DNSs. Left: number of DNSs in each set of simulations (see Table~\ref{tab:ecsim}) as a function of progenitor's metallicity. Right: number of DNSs merging in less than a Hubble time (hereafter: merging DNSs) as a function of progenitor's metallicity. Different runs are indicated by different lines, as explained in the legend and in Table~\ref{tab:ecsim}.}\label{fig:mergingall}
	\end{figure*}	
\subsection{Electron-capture SNe (ECSNe)}
\label{sec:2.1}
NSs can form via CCSN, via ECSN or through the accretion-induced collapse of a white dwarf (WD). In {\sc MOBSE}, the outcome of a CCSN is considered a NS if its mass is less than 3.0 \msun~and a black hole (BH) otherwise. 
This approach is overly simplified, but more constraints on the equation of state of a NS are required for a better choice of the transition between NS and BH.

In the case of both an ECSN and an accretion-induced WD collapse, the NS forms when the degenerate Oxygen-Neon (ONe) core collapses as a consequence of electron-capture reactions, inducing a thermonuclear runaway    \citep{Miyaji1980,  Nomoto1984, Nomoto1987, Nomoto1991, Kitaura2006, Fisher2010, Jones2013, Takahashi2013, Schwab2015, Jones2016}.


In {\sc MOBSE}, we decide whether a star will undergo an ECSN by following the procedure described by \cite{Hurley2000} and \cite{Fryer2012}. First, we look at the Helium core mass at the base of the asymptotic giant branch\footnote{Mass loss during the asymptotic giant branch and dredge-up efficiency are assumed to be the same as in \cite{Hurley2000}.} ($M_{\rm BABG}$). If  $1.6$ \msun~$ \leq M_{\rm BABG} < 2.25$ \msun, the star forms a partially degenerate Carbon-Oxygen (CO) core. If the CO core grows larger than $\sim{}1.08$ M$_\odot$, it can form a degenerate ONe core. If this degenerate ONe core reaches the mass $M_{\rm ECSN}=1.38$ \msun{}, it collapses due to the electron-capture on $^{24}$Mg and on $^{20}$Ne \citep{Miyaji1980,Nomoto1984,Nomoto1987}, otherwise it forms an ONe WD, which can still collapse to a NS if it will accrete sufficient mass.

The outcome of the electron-capture collapse is a NS with baryonic mass $M_{\rm rem, bar} = M_{\rm ECSN}$, which becomes
\begin{equation}
	M_{\rm rem,grav} = \frac{\sqrt{1 + 0.3M_{\rm rem,bar}}-1}{0.15}=1.26 \msun~,
\end{equation}
considering the mass loss due to neutrinos and by using the formula suggested by \citet{Timmes1996}.

Even if only a few per cent of all SN events should be produced by electron-capture reactions in single stars \citep{Poelarends2007,Doherty2015}, this fraction could drastically raise if we consider binary systems \citep{Podsiadlowski2004}.
In binary systems the possibility of accreting material by a companion broadens the mass range of progenitor stars in which the electron-capture collapse may occur, because mass transfer can significantly change the evolution of the core \citep{Sana2012, Dunstall2015,Poelarends2017}. In appendix~\ref{sec:app}, we show that the mass range of ECSNe is crucially affected by binary evolution. In particular, we find that mass transfer tends to widen the mass range of ECSNe.

Recently, \cite{Jones2016} have shown that an ECSN might lead to the ejection of a portion of the degenerate core, rather than to the collapse into a NS. The collapse and the formation of a NS takes place only if the ignition density is $\gtrsim{}2\times{}10^{10}$ g cm$^{-3}$. This finding must be taken into account when interpreting the outcomes of our simulations: our results should be regarded as upper limits to the impact of ECSNe on the statistics of DNSs.

\begin{center}
	\begin{figure*}
		\includegraphics[scale=0.41]{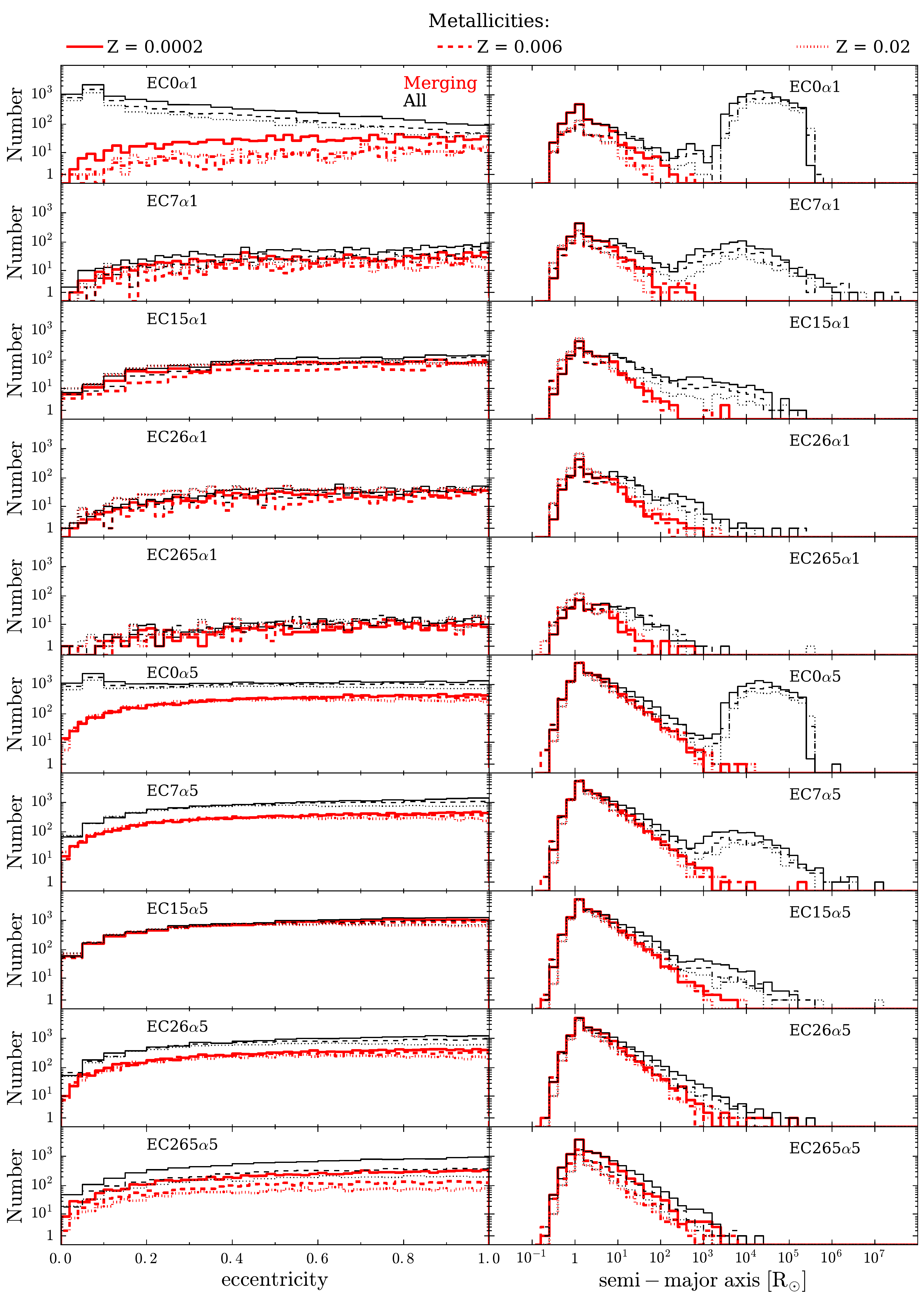}
		\caption{Distribution of eccentricity (left-hand column) and semi-major axis (right-hand column) for all DNSs (black thin lines) and only for merging DNSs (red thick lines). For each simulation we show the distributions obtained at three different metallicities: $Z=0.02$ (dotted lines), 0.006 (dashed lines), and 0.0002 (solid lines). Simulations CC15$\alpha{}1$ and CC15$\alpha{}5$ are not shown in this Figure, because they have already been discussed in Giacobbo \&{} Mapelli (2018).\label{fig:ecca}}
	\end{figure*}	
\end{center}
\begin{center}
	\begin{figure}
		\centering
		\includegraphics[scale=0.3]{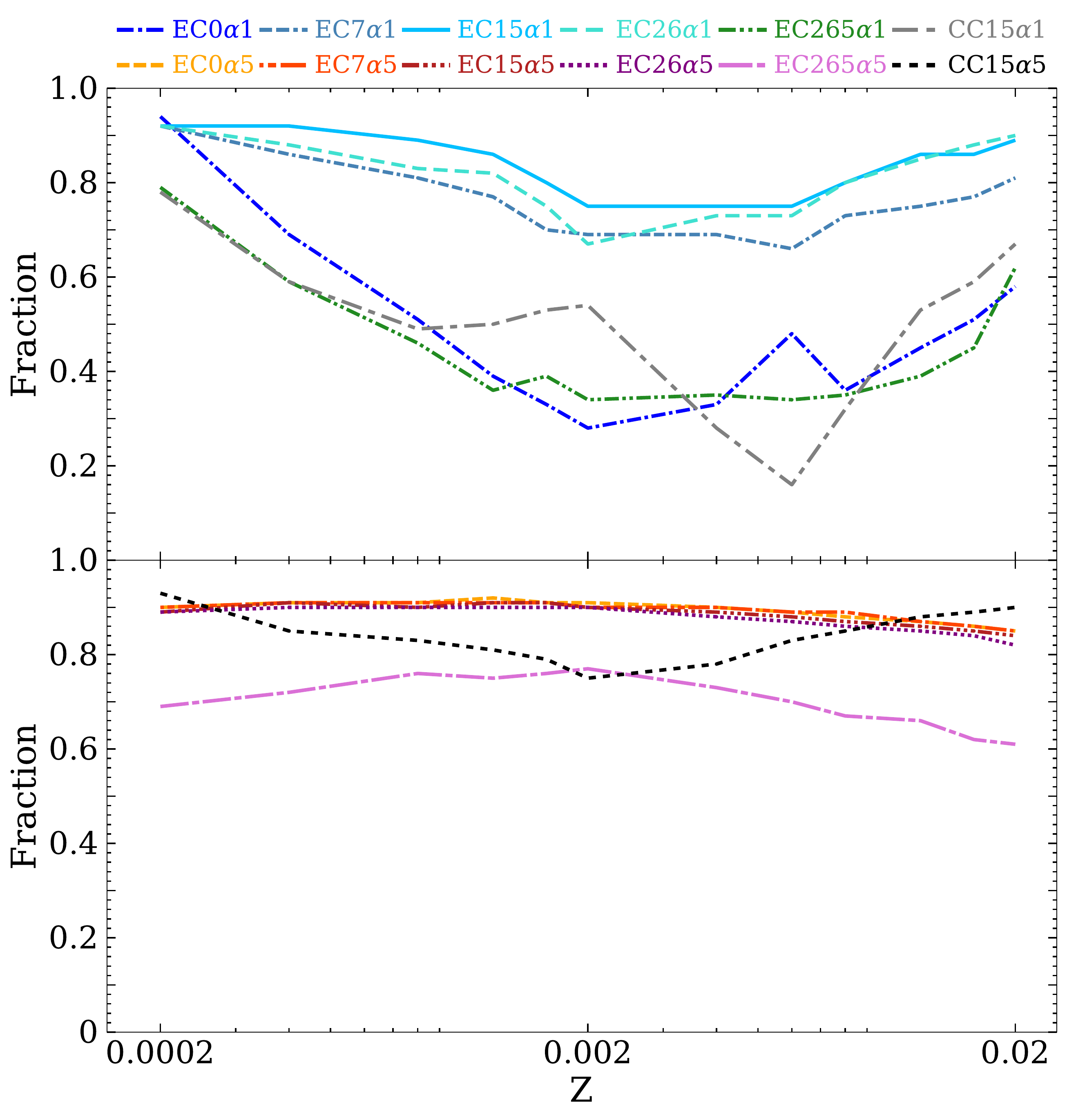}
		\caption{The percentage of merging DNSs which follow the standard scenario (see Sec. 3.2) as a function of progenitor's metallicity. Top: simulations with $\alpha=1$. Bottom: simulations with $\alpha=5$.\label{fig:scenario}}
	\end{figure}	
\end{center}

\subsection{Natal kicks}
\label{sec:2.2}


The natal kick of a NS is drawn from a Maxwellian velocity distribution
\begin{equation}
	f(v,\sigma)=\sqrt{\frac{2}{\pi}}\frac{v^2}{\sigma^3}\exp\left[{-\frac{v^2}{2\sigma^2}}\right] \qquad v \in~ [0,\infty )
\end{equation}
where $\sigma$ is the one dimensional root-mean-square (1D rms) velocity and $v$ is the modulus of the velocity.

Given the uncertainties on the natal kick distribution, we have implemented in {\sc MOBSE} the possibility to draw the natal kick from two Maxwellian curves with a different value of the 1D rms: $\sigma_{\rm CCSN}$ and $\sigma_{\rm ECSN}$, for iron CCSNe and ECSNe, respectively.


$\sigma{} = 265$~km~s$^{-1}$ is adopted as a default value for CCSNe in {\sc MOBSE}. This value was derived by \citet{Hobbs2005}, studying the proper motion of 233 young isolated Galactic pulsars and corresponds to an average natal kick speed of $\sim 420$~km~s$^{-1}$.  

In this paper, we consider different values of $\sigma_{\rm ECSN}$, ranging from 0 to 265~km~s$^{-1}$, to investigate the impact of ECSNe on the statistics of DNSs.

Because low natal kicks might originate not only from ECSNe, but also from  iron CCSNe involving low-mass progenitors and from ultra-stripped SNe, we have run also an extreme case ($\sigma_{\rm CCSN}=\sigma_{\rm ECSN} = 15$~km~s$^{-1}$), in which all NSs receive a low natal kick independently on the SN type (see Table~\ref{tab:ecsim}). We will discuss this extreme case in Section~\ref{sec:alllow}.


\subsection{Simulations and initial distributions}
\label{sec:2.3}
Here we describe the initial conditions used to perform our population-synthesis simulations.   
We randomly draw the mass of the primary star ($m_{\mathrm{1}}$) from a Kroupa initial mass function \citep[IMF,][]{Kroupa2001}
\begin{equation}
	\mathfrak{F}(m_1) ~\propto~ m_1^{-2.3} \qquad \mathrm{with}~~ m_1 \in [5-150]\msun ~.
\end{equation}
The other parameters (mass of the secondary, period and eccentricity), are sampled according to the distributions proposed by \citet{Sana2012}. 
In particular, we obtain the mass of the secondary $m_{\mathrm{2}}$ as follows
\begin{equation}
	\mathfrak{F}(q)~ \propto ~q^{-0.1} \qquad ~~~\mathrm{with}~~~q = \frac{m_2}{m_1}~ \in [0.1-1]~,
\end{equation}
the orbital period $P$ and the eccentricity $e$ from
\begin{equation}
	\mathfrak{F}(\mathscr{P}) ~\propto~ (\mathscr{P})^{-0.55} ~~\mathrm{with}~ \mathscr{P} = \mathrm{log_{10}}(P/\mathrm{day}) \in [0.15-5.5]
\end{equation} 
and 
\begin{equation}
	\mathfrak{F}(e) ~\propto ~e^{-0.42} \qquad ~~\mathrm{with}~~~ 0\leq e < 1~
\end{equation} 
respectively.

For the CE phase we have adopted the $\alpha{}\lambda$ formalism \citep[see][]{Webbink1984,Ivanova2013}. This formalism relies on two parameters, $\lambda{}$ (which measures the concentration of the envelope) and $\alpha{}$ (which quantifies the energy available to unbind the envelope). To compute $\lambda$ we used the prescriptions derived by \citet{Claeys2014} (see their Appendix A for more details) which are based on \citet{Dewi2000}. 

We have run 12 sets of simulations, by changing the value of $\alpha{}$ and that of both $\sigma_{\rm ECSN}$ and $\sigma_{\rm CCSN}$ (see Table~\ref{tab:ecsim}).

  In the first 10 simulations reported in Table~\ref{tab:ecsim}, we have fixed $\sigma_{\rm CCSN}=265$~km~s$^{-1}$ and we have varied $\alpha{}=1,\,{}5$ and $\sigma_{\rm ECSN}=0,7,15,26,265$~km~s$^{-1}$ (corresponding to an average natal kick of about $0,11,23,41,420$~km~s$^{-1}$, respectively).

  In the last two simulations reported in Table~\ref{tab:ecsim} (CC15$\alpha{}1$ and CC15$\alpha{}5$), we have set $\sigma_{\rm CCSN} = \sigma_{\rm ECSN} = 15$~km~s$^{-1}$ for both $\alpha{}=1,5$. We will discuss simulations CC15$\alpha{}1$ and CC15$\alpha{}5$ in Section~\ref{sec:alllow}, while in the following sections we will focus on the other 10 simulations (i.e. on the effect of $\sigma_{\rm ECSN}$ on the statistics of DNSs).

 Finally, for each set of simulations we considered 12 sub-sets with different metallicities $Z=0.0002$, $0.0004$, $0.0008$, $0.0012$, $0.0016$, $0.002$, $0.004$, $0.006$, $0.008$, $0.012$, $0.016$ and $0.02$. In each sub-set, we simulated $10^7$ binary systems. Thus, each of sets of simulations is composed of $1.2\times10^8$ massive binaries. 

\section{Results}
\begin{center}
	\begin{figure*}
		\includegraphics[scale=0.4]{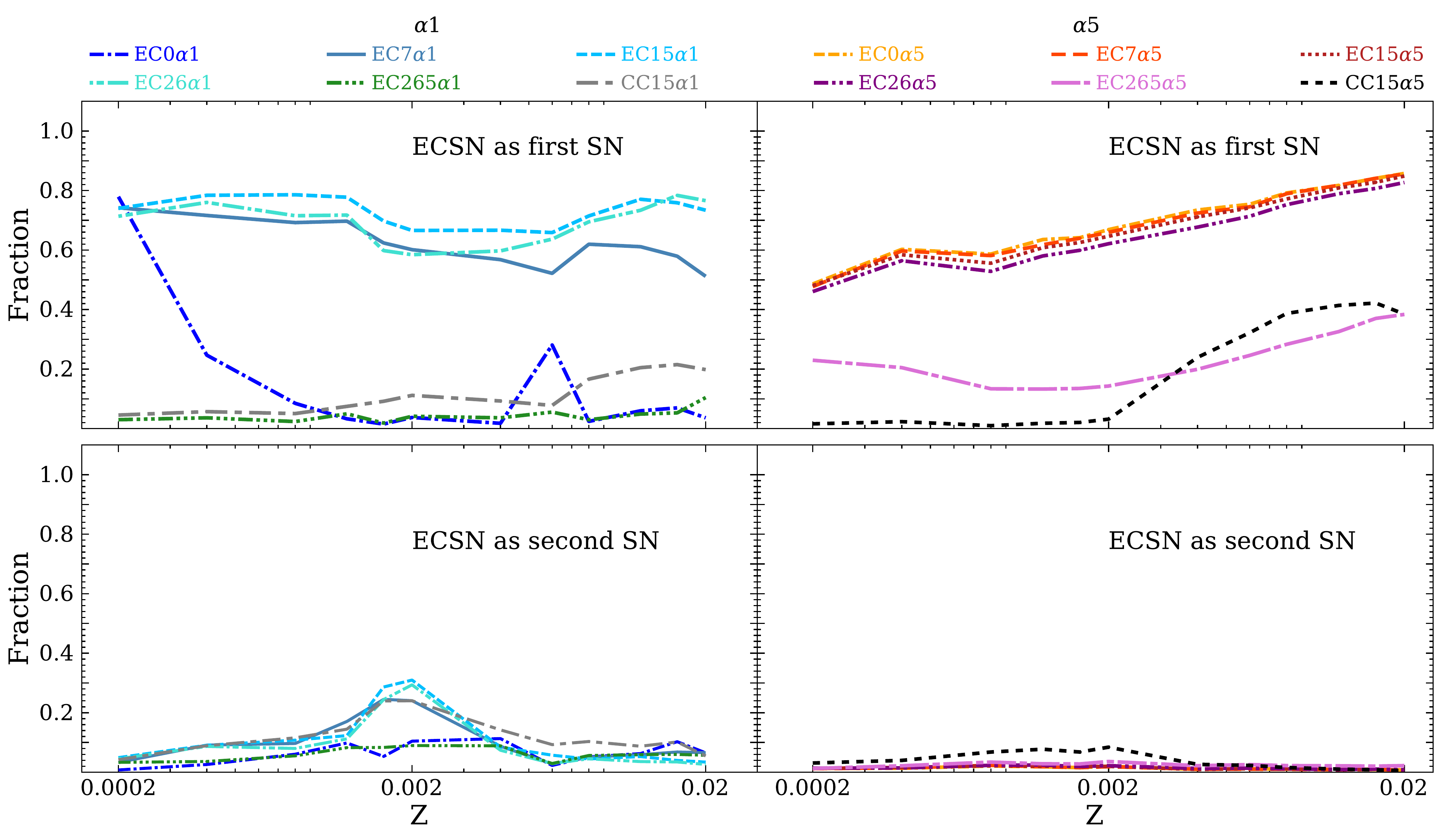}
		\caption{\label{fig:fraction}Top (bottom) panels: fraction of merging DNSs in which the first (second) SN is an ECSN as a function of progenitor's metallicity. Left-hand (right-hand) panels: simulations with $\alpha{}=1$ ($\alpha{}=5$).}
	\end{figure*}	
\end{center}
\subsection{Impact of $\sigma{}_{\rm ECSN}$ on DNSs}
\label{sec:4}
The left-hand panel of Figure~\ref{fig:mergingall} shows all DNSs formed in our simulations as a function of metallicity.  It is apparent that the lower $\sigma_{\rm ECSN}$ is, the higher the total number of DNSs. This is not surprising, because a lower $\sigma_{\rm ECSN}$ implies a lower probability to unbind the system.

This effect is particularly strong for the simulations with $\alpha{}=1$, in which the number of DNSs is $\sim{}10-25$ times higher if $\sigma{}_{\rm ECSN}=0$ than if  $\sigma{}_{\rm ECSN}=265$~km~s$^{-1}$. In the simulations with $\alpha{}=5$, the number of DNSs is $3-6$ times higher if $\sigma{}_{\rm ECSN}=0$ than if  $\sigma{}_{\rm ECSN}=265$~km~s$^{-1}$. We also note that DNSs form more efficiently if $\alpha{}=5$ than if $\alpha{}=1$.



In our simulations, the number of DNSs depends on metallicity, especially if $\alpha{}=1$. In particular, the number of DNSs is minimum for $Z\sim 0.002$. This trend originates from the evolution of stellar radii of $\sim{}8-20$ M$_\odot$ stellar progenitors, which are significantly larger for $Z\sim{}0.002$, than for the other metallicities (especially in the terminal main sequence and in the HG phases). The trend is stronger for $\alpha{}=1$ than for $\alpha{}=5$, because a low value of $\alpha{}$ corresponds to a more efficient shrinkage of the orbit during CE: two main sequence or HG stars are more likely to merge during CE if $\alpha{}$ is low.

The right-hand panel of Figure~\ref{fig:mergingall} shows only the DNSs which merge in less than a Hubble time (hereafter: merging DNSs). In the simulations with $\alpha{}=5$, we find again a monotonic trend with $\sigma_{\rm ECSN}$, but the differences are much less significant.

In the simulations with $\alpha=1$ the number of merging DNSs does not show a monotonic trend with $\sigma{}_{\rm ECSN}$: runs with $\sigma{}_{\rm ECSN}=7-26$~km~s$^{-1}$ produce a factor of $\sim{}5$ more merging DNSs than simulations with $\sigma{}_{\rm ECSN}=0$ and 265~km~s$^{-1}$. The only exception is represented by very metal-poor stars ($Z=0.0002$), for which the number of merging DNSs with $\sigma{}_{\rm ECSN}=0$ is similar to the one of systems with $\sigma{}_{\rm ECSN}=7-26$~km~s$^{-1}$. 

This behavior can be easily explained by considering that the merging time ($t_{\rm gw}$) due to GW emission depends on both the eccentricity ($e$) and the semi-major axis ($a$) as \citep{Peters1964}
\begin{equation}\label{eq:eqpeters}
	t_{\rm gw} = \frac{5}{256}\frac{c^5}{G^3} \frac{a^4(1-e^2)^{7/2}}{m_1m_2(m_1+m_2)}~,
\end{equation}
where $c$ is the speed of light, $G$ is the gravitational constant, and $m_1$ ($m_2$) is the mass of the primary (secondary) member of the binary.

Equation~\ref{eq:eqpeters} implies that more eccentric binaries have a shorter merging time. Moderate natal kicks do not unbind a binary, but increase its eccentricity, shortening its merging time.
Since most binaries evolve through processes which tend to circularize their orbits (e.g. tidal torques, mass transfer and CE phase), the natal kicks are a fundamental ingredient to obtain highly eccentric orbits.


This behavior is shown in the left-hand column of Figure~\ref{fig:ecca}, where the initial eccentricity distribution of all DNSs is compared with that of the merging DNSs (here ``initial'' refers to the time when the second NS is formed). A large number of DNSs have initial eccentricity close to zero in run~EC0$\alpha1$ (corresponding to $\sigma_{\rm ECSN}=0$ and $\alpha{}=1$), but only very few of them merge within a Hubble time. 

Many DNSs have initial eccentricity close to zero and most of them do not merge within a Hubble time also in run~EC0$\alpha5$ (corresponding to $\sigma_{\rm ECSN}=0$ and $\alpha{}=5$). However, run~EC0$\alpha5$ is also efficient in producing DNSs with non-zero eccentricity, which are able to merge within a Hubble time. In contrast, only few DNSs with eccentricity close to zero form in the other runs, because of the SN kicks. We note that the second NS originates from an ECSN in the vast majority of DNSs with eccentricity $e\sim{}0$.


The right-hand column of Figure~\ref{fig:ecca} compares the distribution of the initial semi-major axis of all DNSs with that of the merging systems. We see that increasing $\sigma_{\rm ECSN}$ the widest systems tend to disappear, because they can be disrupted more easily by the natal kicks.

\subsection{DNS formation channels}
\label{sec:formation}
From our simulations we find that the most likely formation channel for merging DNSs is consistent with the standard scenario described in \citet{Tauris2017} (see their Figure 1): first the primary star expands and fills its  Roche lobe, transferring mass to the companion; then the primary explodes leaving a NS; when the secondary expands, the system enters CE; after CE ejection, the system is composed of a NS and a naked Helium star and the NS starts stripping its companion; the stripped Helium star undergoes a SN explosion, which is most likely an ultra-stripped SN \citep{Tauris2013,Tauris2015,Tauris2017}; the final system is a close DNS which will merge within a Hubble time.  

Figure~\ref{fig:scenario} shows the fraction of merging DNSs which follow the standard scenario we have just described ($f_{\rm std}$). For $\alpha=5$, $f_{\rm std}$ is nearly independent of the metallicity of the progenitor, while it depends on the natal kicks. At low kicks ($\sigma_{\rm ECSN}\leq 26$~km~s$^{-1}$) $>>80$ per cent of  merging DNSs form via the standard scenario, while if $\sigma_{\rm ECSN}= 265$~km~s$^{-1}$ the percentage lowers to $\sim{}60-70$ per cent.

For $\alpha=1$, $f_{\rm std}$ depends on both the metallicity and the natal kicks. For a given kick distribution, $f_{\rm std}$ is minimum at metallicity $Z\sim 0.0016 - 0.006$ (especially in run EC0$\alpha1$ and EC265$\alpha1$), while for a fixed metallicity $f_{\rm std}$ is maximum ($\sim 80 - 90$ per cent) for $\sigma{}_{\rm ECSN}=7-26$~km~s$^{-1}$.

This behavior confirms that ECSNe are a fundamental process for the formation of DNSs, but what is the fraction of systems undergoing an ECSN? Is ECSN more frequently the first or the second SN of a merging system?

Figure~\ref{fig:fraction} shows the fraction of merging DNSs in which at least one of the two SN explosions is an ECSN. 
Most merging DNSs ($\sim{}50-90$ per cent) undergo at least one ECSN in the vast majority of simulations (EC7$\alpha{}1$, EC15$\alpha{}1$, EC26$\alpha{}1$, EC0$\alpha{}5$, EC7$\alpha{}5$, EC15$\alpha{}5$ and EC26$\alpha{}5$). In the simulation EC0$\alpha{}1$ ($\sigma{}_{\rm ECSN}=0$ and $\alpha{}=1$), ECSNe are important at low metallicity ($Z=0.0002$) and negligible for intermediate and high metallicity. Only in the simulations with large ECSN kicks (runs EC265$\alpha{}1$ and EC265$\alpha{}5$), the fraction of DNSs undergoing at least one ECSN is always less than 50 per cent.

Moreover, in simulations with $\alpha=5$ the percentage of DNSs which undergo at least one ECSN increases with the progenitor's metallicity.


Overall, we find that the ECSN is the first SN in the vast majority of merging DNSs. Less than $\sim{}10$ per cent of merging DNSs go through an ECSN as second SN, independently of the assumptions about natal kicks and CE efficiency. This result is in agreement with \cite{Chruslinska2017} and \cite{Kruckow2018} (but see \citealt{Tauris2017} for a different argument).

This is likely due to the fact that the first SN explosion occurs before that other processes (e.g. a CE phase) are able to shrink the binary; therefore the system is less bound and it can be more easily disrupted if the natal kick of the newborn NS is too strong. In contrast, the second SN explosion tends to occur after a CE, when the system is usually on a very close and less eccentric orbit, hence it can survive even stronger kicks. Moreover, the fact that the second SN explosion induces a high kick velocity facilitates the formation of highly eccentric orbits, which are more likely to merge via GW emission.

The fact that the ECSN is often the first SN occurring in a merging DNS might seem awkward, because ECSN progenitors are usually less massive than iron CCSN progenitors. Indeed, this happens because most merging DNSs originate from very close binary systems, in which the primary has lost a significant fraction of its mass by mass transfer during Roche lobe overflow. Because of mass loss, the primary enters the regime of ECSNe. In contrast, the secondary accretes part of the mass lost by the primary and enters the regime of iron CCSNe. This explains why the first SN is more often an ECSN in the progenitors of merging DNSs.


\subsection{GW170817-like systems}
\begin{center}
	\begin{figure}
		\includegraphics[scale=0.36]{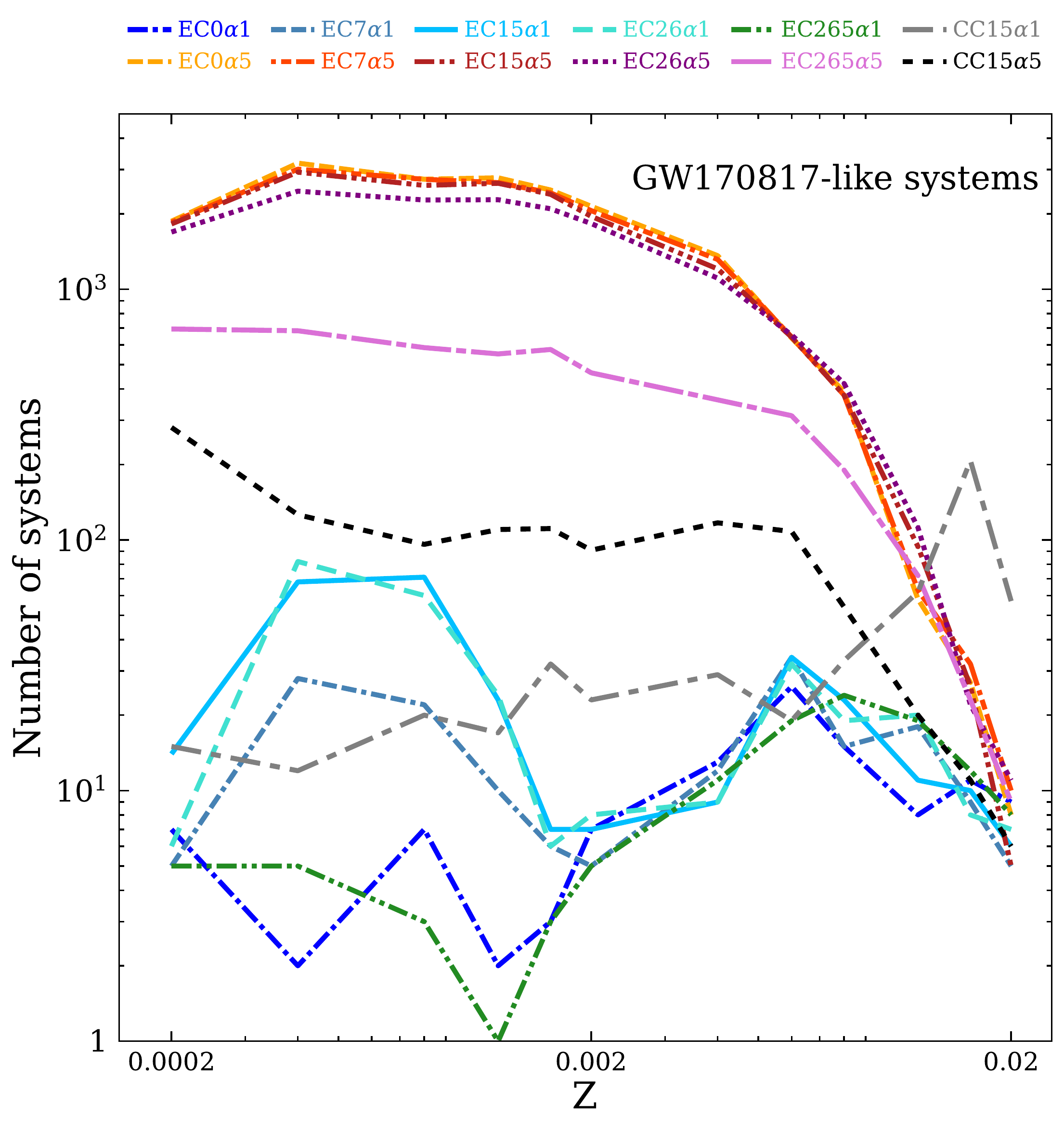}
		\caption{Number of GW170817-like simulated merging DNSs as a function of progenitor's metallicity.\label{fig:gwlike}}
	\end{figure}	
\end{center}

Figure~\ref{fig:gwlike} shows the number of GW170817-like systems that form in our simulations. We define as GW170817-like systems those merging DNSs with $M_{\rm rem,1} \in [1.36, 1.60] \msun$ and $M_{\rm rem,2} \in [1.17, 1.36] \msun$ ($M_{\rm rem,1}$ and $M_{\rm rem,2}$ being the mass of the primary and of the secondary NS, assuming effective spin $\leq{}0.05$, \citealt{Abbott2017b}). Because of its large mass ($1.36 - 1.60 \msun$), the most massive component of GW170817-like systems cannot have formed via ECSN. In other words, at least one of the two SNe must be a CCSN, in order to form a GW170817-like system.

Figure~\ref{fig:gwlike} shows that at high metallicity ($Z\gtrsim{}0.002$ for $\alpha{}=1$ and $Z\gtrsim{}0.012$ for $\alpha{}=5$) all simulations follow a similar trend independently of the value of $\sigma_{\rm ECSN}$, while for lower metallicities the number of GW170817-like systems becomes sensitive to the value of $\sigma_{\rm ECSN}$. In particular, the higher $\sigma_{\rm ECSN}$ is, the lower the number of GW170817-like systems. Furthermore, in the simulations with $\alpha=5$ the number of GW170817-like systems increases with decreasing metallicity.

\begin{center}
	\begin{figure*}
		\includegraphics[scale=0.4]{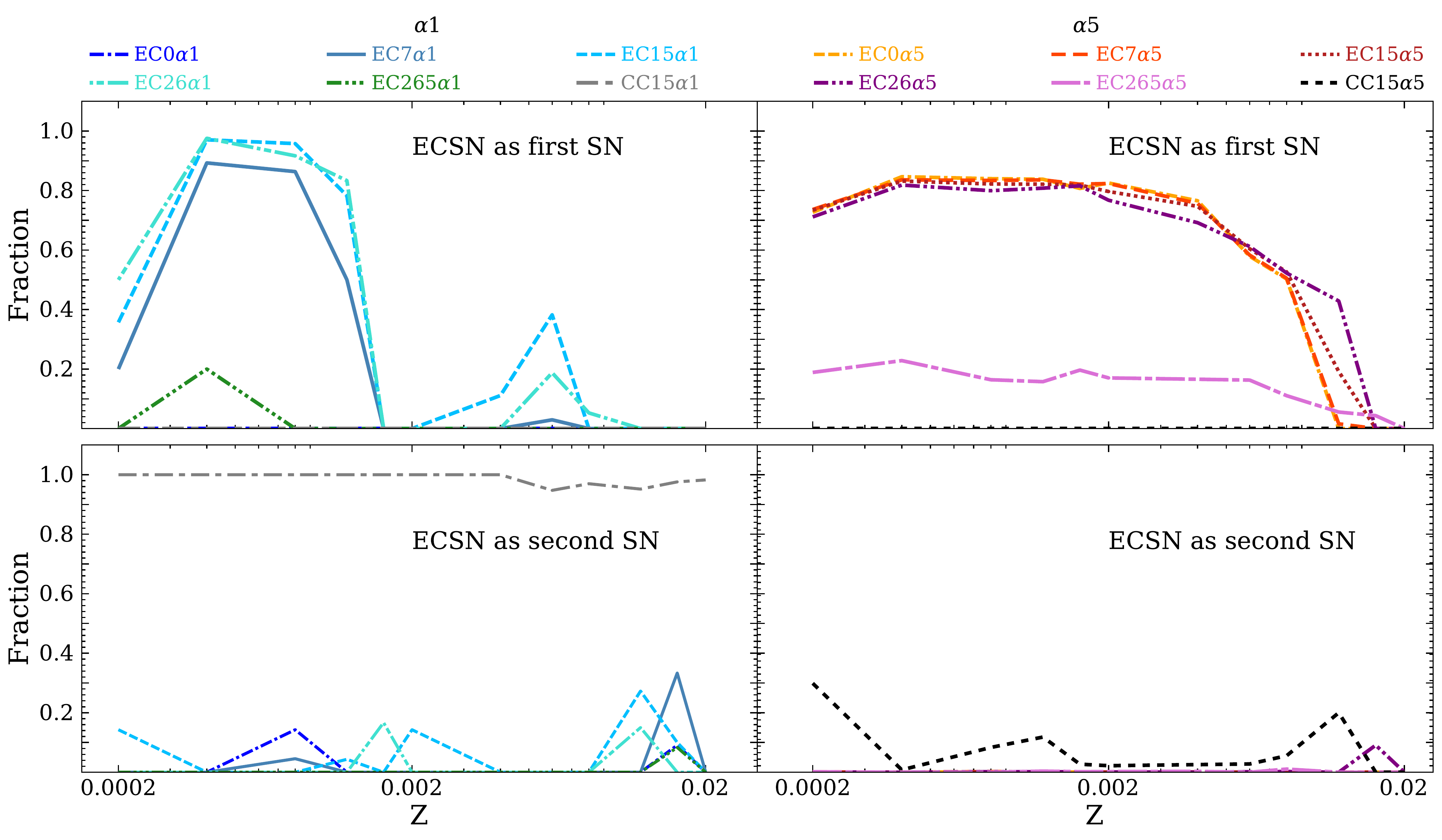}
		\caption{Top (Bottom) panels: fraction of GW170817-like systems in which the first (second) SN is an ECSN as a function of progenitor's metallicity. The left-hand (right-hand) panels are for the simulation with $\alpha=1$ ($\alpha=5$). \label{fig:fractiongw}}
	\end{figure*}	
\end{center}

The reason is that at high metallicity the majority of GW170817-like systems form from binaries which undergo two iron CCSNe (see Figure~\ref{fig:fractiongw}), while at low metallicity most of the progenitors pass through at least one ECSN. Figure~\ref{fig:fractiongw} shows that the effect of increasing the value of $\alpha$ is to increase the maximum metallicity at which the majority of GW170817-like systems form through at least one ECSN from $Z\sim 0.002$ ($\alpha=1$) to $Z \sim 0.012$ ($\alpha=5$).

\subsection{Low kicks in iron core-collapse SNe}\label{sec:alllow}
Low natal kicks might occur not only in ECSNe but also in iron CCSNe, especially in the case of low-mass progenitors \citep{Mueller2016} and ultra-stripped SNe \citep{Tauris2017}. Moreover, \cite{Giacobbo2018b} and \cite{Mapelli2018} have shown that population-synthesis simulations can reproduce the local merger rate of DNSs inferred from GW170817 \citep{Abbott2017b} only if low natal kicks ($\lesssim{}50$~km~s$^{-1}$) are assumed for all DNSs.

Thus, in this Section we discuss how much the main results of this paper change if we make the extreme assumption that all NSs receive a low natal kick, i.e. $\sigma{}_{\rm ECSN}=\sigma{}_{\rm CCSN}=15$~km~s$^{-1}$. To this purpose, we have considered two additional runs: CC15$\alpha{}1$ and CC15$\alpha{}5$, which have been already presented in \cite{Giacobbo2018b} and \cite{Mapelli2018}. Simulation CC15$\alpha{}1$ (CC15$\alpha{}5$) differs from simulation EC15$\alpha{}1$ (EC15$\alpha{}5$) only for the choice of $\sigma{}_{\rm CCSN}$ (see Table~\ref{tab:ecsim}).

From Figure~\ref{fig:mergingall} it is apparent that the number of DNSs  is about one order of magnitude larger in simulation CC15$\alpha{}1$ (CC15$\alpha{}5$) than in simulation EC15$\alpha{}1$ (EC15$\alpha{}5$). This is not surprising, because a lower CCSN kick implies a lower probability to unbind the system.

On the other hand, if we consider only the DNSs merging within a Hubble time, we find an interesting difference. The number of merging DNSs is a factor of ten larger in simulation CC15$\alpha{}1$ than in simulation EC15$\alpha{}1$, whereas the number of merging DNSs in simulations CC15$\alpha{}5$ and EC15$\alpha{}5$ are comparable. Moreover, simulations CC15$\alpha{}5$ and EC15$\alpha{}5$ show a significantly different trend with metallicity. As already discussed in \cite{Giacobbo2018b}, there is a strong interplay between the effects of natal kicks and those of CE efficiency.

Figure~\ref{fig:fraction} shows that the percentage of merging DNSs which underwent at least one ECSN is dramatically affected by the choice of $\sigma{}_{\rm CCSN}$: less than $\sim{}40$ per cent of all merging  DNSs underwent at least one ECSN if $\sigma{}_{\rm CCSN}=15$~km~s$^{-1}$. This difference is particularly strong for the first SN. In fact, the binary system is still quite large at the time of the first SN explosion and can be easily broken by the SN kick.

This result has relevant implications for the GW170817-like systems. As shown in Fig.~\ref{fig:fractiongw}, no GW170817-like systems underwent an ECSN as first SN in simulations CC15$\alpha{}1$ and CC15$\alpha{}5$. This comes from the trend we described above, plus the fact that the first SN usually produces the most massive NS of the system in runs CC15$\alpha{}1$ and CC15$\alpha{}5$. In our models, the mass of a NS born from ECSN is assumed to be 1.26 M$_\odot{}$, insufficient to match the mass of the primary member of GW170817 (under the assumption of low spins). In contrast, the second SN is always an ECSN in the GW170817-like systems formed in simulation CC15$\alpha{}1$. We stress, however, that this result critically depends on the assumption about the mass of a NS formed via ECSN \citep{Hurley2000,Fryer2012}.

\section{Summary}
We have investigated the importance of ECSNe on the formation of DNSs. ECSNe are thought to occur frequently in interacting binaries \citep{Podsiadlowski2004,Tauris2017} and to produce relatively small natal kicks \citep{Dessart2006,Jones2013,Schwab2015}. We assumed that natal kicks generated by ECSNe (iron CCSNe) are distributed according to a Maxwellian function with 1D rms $\sigma_{\rm ECSN}$ ($\sigma_{\rm CCSN}$). 

First, we have assumed $\sigma{}_{\rm CCSN}=265$~km~s$^{-1}$ (according to \citealt{Hobbs2005}) and we have explored five different values of $\sigma{}_{\rm ECSN}=0,$ 7, 15, 26 and 265~km~s$^{-1}$.  
Moreover, we have also investigated the impact of CE, by considering $\alpha{}=1$ and $\alpha{}=5$.

We find that the number of simulated DNSs scales inversely with $\sigma_{\rm ECSN}$. In particular, the largest (smallest) number of DNSs form if  $\sigma{}_{\rm ECSN}=0$ ($\sigma{}_{\rm ECSN}=265$~km~s$^{-1}$). This effect is maximum for $\alpha{}=1$, while it is only mild for $\alpha{}=5$. 

The number of DNSs merging within a Hubble time also depends on $\sigma_{\rm ECSN}$, but with a rather different trend depending on the assumed value for $\alpha$. For $\alpha=5$, the number of merging systems follows the same trend as the total number of DNSs. For $\alpha=1$ the number of DNS mergers is maximum for $\sigma{}_{\rm ECSN}=7-26$~km~s$^{-1}$, while it drops by a factor of $\sim{}3-10$ if $\sigma_{\rm ECSN}=0$ and if $\sigma_{\rm ECSN}=265$~km~s$^{-1}$.

The reason is that very large kicks ($\sigma_{\rm ECSN}=265$~km~s$^{-1}$) completely break the binary, while moderate kicks ($\sigma{}_{\rm ECSN}=7-26$~km~s$^{-1}$) leave the binary bound but increase its eccentricity. A larger eccentricity implies a shorter timescale for merger by GW emission, as shown by \cite{Peters1964}. In contrast, null natal kicks produce a large number of systems with zero initial eccentricity, which have longer merger times.

A large percentage ($\sim{}50-90$ per cent) of merging DNSs undergo at least one ECSN explosion in most of our simulations. This percentage drops below 40 per cent only if $\sigma{}_{\rm ECSN}=265$~km~s$^{-1}$, or if $\sigma{}_{\rm CCSN}=15$~km~s$^{-1}$, or if $\sigma{}_{\rm ECSN}=0$~km~s$^{-1}$, $\alpha{}=1$ and $Z>0.0002$.

In the majority of merging DNSs, the ECSN is the first SN occurring in the binary. This happens because, in most cases, the first SN occurs before the binary has shrunk significantly (e.g. by CE) and is easily broken if the kick is too strong. 



Moreover, we have selected the simulated DNSs whose mass matches that of GW170817. We call these systems GW170817-like systems. At high metallicity ($Z \gtrsim 0.002$ for $\alpha=1$ and $Z \gtrsim 0.012$ for $\alpha=5$) the formation of GW170817-like systems is independent of $\sigma{}_{\rm ECSN}$, because most GW170817-like systems form through iron CCSNe, while for lower metallicity most GW170817-like systems undergo at least one ECSN and their statistics depends on $\sigma{}_{\rm ECSN}$.

Finally, we have considered an extreme case in which not only ECSNe but also CCSNe are associated to low kicks, by imposing $\sigma{}_{\rm CCSN}=\sigma_{\rm ECSN}=15$~km~s$^{-1}$. \cite{Mapelli2018} and \cite{Giacobbo2018b} suggest that this extreme assumption is necessary to match the local DNS merger rate density inferred from GW170817 \citep{Abbott2017b}.

  The number of simulated DNSs increases by a factor of ten if we assume $\sigma{}_{\rm CCSN}=15$~km~s$^{-1}$, because less binary systems are disrupted by the first SN explosion. Moreover, this assumption strongly suppresses the percentage of merging DNSs (especially GW170817-like systems) which evolved through an ECSN as first SN. 
  These results confirm the importance of natal kicks to understand the properties of merging DNSs.

\section*{Acknowledgements}
We thank the referee, Samuel Jones, for his critical reading which significantly improved this work. The authors are grateful to Alessandro Bressan, Mario Spera and Chris Pankow for useful discussions. Numerical calculations have been performed through a CINECA-INFN agreement and through a CINECA-INAF agreement, providing access to resources on GALILEO and MARCONI at CINECA. 
 NG acknowledges financial support from Fondazione Ing. Aldo Gini and thanks the Institute for Astrophysics and Particle Physics of the University of Innsbruck for hosting him during the preparation of this paper.
 MM  acknowledges financial support from the MERAC Foundation through grant `The physics of gas and protoplanetary discs in the Galactic centre', from INAF through PRIN-SKA `Opening a new era in pulsars and compact objects science with MeerKat', from MIUR through Progetto Premiale 'FIGARO' (Fostering Italian Leadership in the Field of Gravitational Wave Astrophysics) and 'MITiC' (MIning The Cosmos: Big Data and Innovative Italian Technology for Frontier Astrophysics and Cosmology), and from the Austrian National Science Foundation through FWF stand-alone grant P31154-N27 `Unraveling merging neutron stars and black hole - neutron star binaries with population-synthesis simulations'. 
 This work benefited from support by the International Space Science Institute (ISSI), Bern, Switzerland, through its International Team programme ref. no. 393
{\it The Evolution of Rich Stellar Populations \& BH Binaries} (2017-18).




\bibliographystyle{mnras}
\bibliography{biblio} 


\appendix{}
\section{ECSN in binary evolution}
\label{sec:app}
\begin{figure*}
  \centering
  \includegraphics[width=0.76\textwidth]{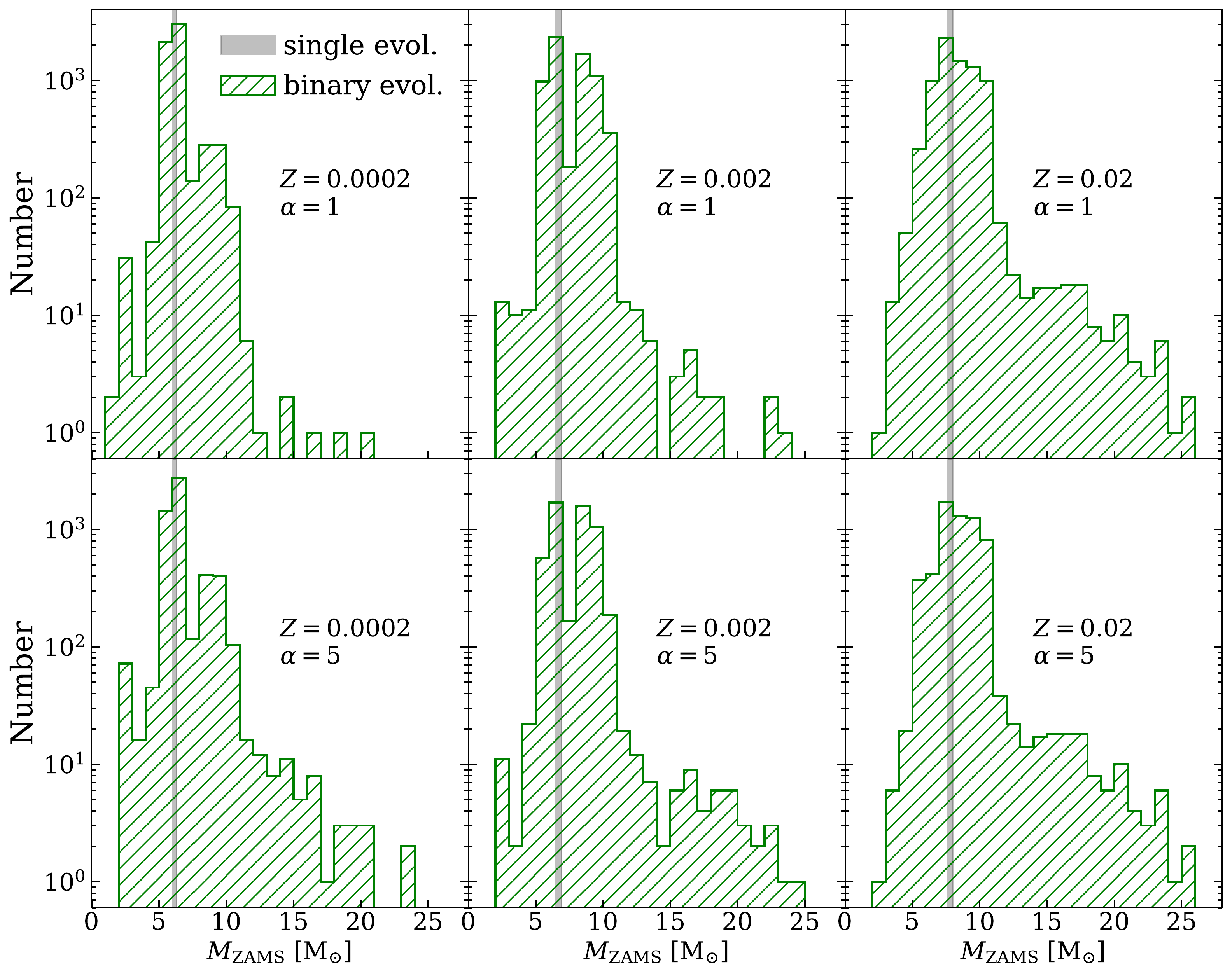}
  \caption{Green hatched histograms: distribution of the zero-age main sequence mass ($M_{\rm ZAMS}$) of binary stars undergoing an ECSN. From left to right: $Z=0.0002,0.002,$ and 0.02, respectively.  For each metallicity, we consider $10^5$ binary systems randomly selected from runs EC15$\alpha{}1$ (top) and EC15$\alpha{}5$ (bottom). Grey shadowed regions (with arbitrary normalization): mass range of single stars undergoing an ECSN. \label{fig:app1}}
\end{figure*}
We describe the impact of binary evolution on the mass range of stars which undergo an ECSN. We consider only the case with $\sigma_{\rm ECSN} = 15$ km s$^{-1}$ and $\sigma_{\rm CCSN} = 265$ km s$^{-1}$, because  different assumptions on natal kicks do not affect the mass range. Thus, we analyze a sub-sample of runs EC15$\alpha{}1$ and EC15$\alpha{}5$ with metallicity  $Z=0.02, 0.002, 0.0002$ (we randomly select $10^5$ binaries for each metallicity). 

In Fig.~\ref{fig:app1}, we compare the zero-age main sequence (ZAMS) mass range of single stars which undergo an ECSN (gray regions) with the ZAMS mass distribution of stars which undergo an ECSN as a consequence of binary evolution (green histograms).  
In agreement with previous studies \citep[e.g.][]{Podsiadlowski2004,Sana2012,Dunstall2015,Poelarends2017}, we find that binary evolution broadens the mass range of stars undergoing an ECSN.

This happens because non-conservative mass transfer changes the mass of close binary members significantly. For instance, it can happen that a primary star with $M_{\rm ZAMS}$ up to $\sim 25$ \msun~loses most of its mass and enters the ECSN regime. On the other hand, if a secondary star with $M_{\rm ZAMS} \gtrsim 2.5$ \msun~accretes enough matter from the companion it can even undergo an ECSN. 

The ZAMS mass range of stars undergoing ECSNe seems to be almost insensitive on the efficiency of CE, especially at high metallicity. Furthermore, the mass range mildly depends on metallicity: the lower the metallicity is, the lower the maximum mass for a star to undergo an ECSN as a consequence of mass transfer (this is likely an effect of stellar winds). 

Finally, we estimate that $\sim{}16$ per cent of NSs undergo an ECSN if $\alpha=1$ and $Z=0.02$, when we account for binary evolution. This percentage increases with decreasing metallicity. In particular, we find that $\sim 13$ per cent of SNe are ECSNe if $\alpha{}=1$ and $Z=0.002$, and $\sim 8$ per cent if $\alpha{}=1$ and $Z=0.0002$). If $\alpha=5$, we find a slightly lower number of ECSNe: $\sim 13,10,7$ per cent at $Z=0.02, 0.002, 0.0002$, respectively.




\bsp	
\label{lastpage}
\end{document}